# Brief Report

neuroscience

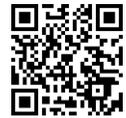

# Control over stress induces plasticity of individual prefrontal cortical neurons: A conductance-based neural simulation

Juan A. Varela[1], Jungang Wang[1], Andrew L. Varnell[1] & Donald C. Cooper[1]*

**Behavioral control over stressful stimuli induces resilience to future conditions when control is lacking. The medial prefrontal cortex(mPFC) is a critically important brain region required for plasticity of stress resilience. We found that control over stress induces plasticity of the intrinsic voltage-gated conductances of pyramidal neurons in the PFC. To gain insight into the underlying biophysical mechanisms of this plasticity we used the conductance-based neural simulation software tool, NEURON, to model the increase in membrane excitability associated with resilience to stress. A ball and stick multicompartment conductance-based model was used to realistically fit passive and active data traces from prototypical pyramidal neurons in neurons in rats with control over tail shock stress and those lacking control. The results indicate that the plasticity of membrane excitability associated with control over stress can be attributed to an increase in $Na^+$ and $Ca^{2+}$ T-type conductances and an increase in the leak conductance. Using simulated dendritic synaptic inputs we observed an increase in excitatory postsynaptic summation and amplification resulting in elevated action potential output. This realistic simulation suggests that control over stress enhances the output of the PFC and offers specific testable hypotheses to guide future electrophysiological mechanistic studies in animal models of resilience and vulnerability to stress.**

Traumatic and stressful life events affect individuals differently depending on the degree of control they have over the stressors. An individual's sense of actual or perceived control over stressors is regarded as a potent coping mechanism that can boost resilience to a number of psychiatric conditions like anxiety, depression and posttraumatic stress disorder (PTSD). In rodent models, behavioral control over a stressor like mild electric tail shock prevents the physiological and learning deficits resulting from lack of control[1,2]. It has been demonstrated that the underlying mechanism of this type of control-mediated resilience requires lasting changes in neurotransmission in the medial prefrontal cortex (mPFC)[3].

We have recent unpublished data from electrophysiological recordings in brain slices of rats allowed to escape tail shock stress that shows pyramidal neuronal plasticity in the deep layers of the pre-limbic subregion of the mPFC. Compared to pyramidal neurons obtained from rats that lack control over shock or home caged, no-shock group the rats that are allowed to learn to escape undergo plasticity in their pyramidal neurons that increases their action potential output. As a group, rats with control have pyramidal neurons that exhibit a faster membrane time constant (TC), larger action potential (AP) amplitude, faster AP rise rate and a larger post-spike after-depolarization (ADP).

Because of the dynamical voltage, time and morphological interactions involved in the integration of excitatory and inhibitory input to pyramidal neurons we utilized a simplified conductance-based modeling approach to determine which critical ion channels could be modified to produce the physiological findings we observed. To accomplish this we utilized the neural simulation tool NEURON to examine the relationships between various conductances to fit the measure physiological data, and determine the conductance changes between the different groups. The utility of this approach is that we are able to refine a model that allows hypothesis testing for future electrophysiological experiments to identify the précised cellular mechanisms of plasticity associated with control over stress.

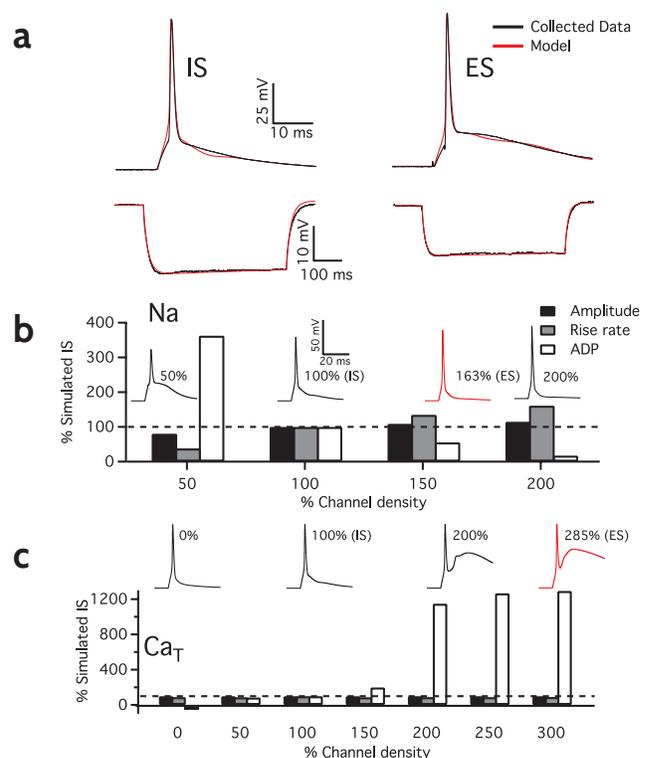

Fig 1. *Modeling escapable and inescapable conditions.* **a.** The data (black) was first fitted (red) by setting the passive parameters using a hyperpolaraizing current pulse (bottom) and then the active parameters using the action potential trace (top). Left, fit for the IS condition, right, fit for the ES condition. **b.** The only active parameters that needed to be changed to account for the IS/ES differences (AP amplitude, AP rise rate and ADP) were the Na (top) and the T-type $Ca^{2+}$ (bottom) channel densities. Top, increasing the Na channel density increases the AP amplitude (black bars) and rise rate (grey bars) but decreases the ADP (white bars), the ES model (red) has 63% more Na channels compared to IS model (100%). **c.** increasing the Ca channel density dramatically increases the ADP (white bars) while the AP amplitude (black bars) remains the same, the ES model (red) has 185% more T-type $Ca^{2+}$ channels than the IS model (100%).

## RESULTS

To account for the difference between the escapable shock (ES) and inescapable shock (IS) groups we modeled representative traces from each group (Fig. 1). The differences are a reflection of changes in the biophysical properties of the neurons[4]. We developed a simple ball and stick model and fitted the data to elucidate the underlying conductances.

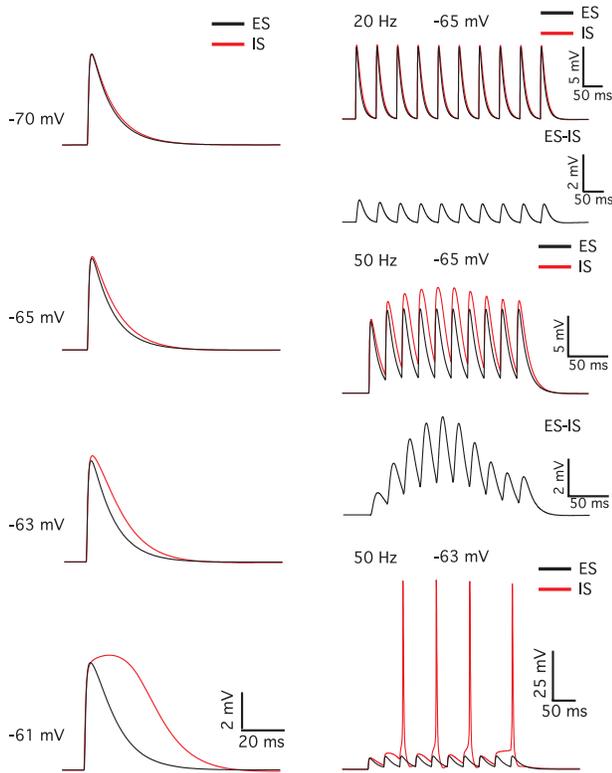

Fig 2. *EPSPs are functionally enhanced in escapable conditions compared to inescapable.* The same simulated synaptic input was delivered while changing the holding potential to more depolarizing values for both the ES (red) and IS (black) model. The model predicts a larger effect of the synaptic boosting in ES compared to IS. Synaptic summation. Trains of the same 10 synaptic input were delivered at different frequencies for both the ES (red) and IS (black) models at the same holding potential, synaptic inputs in ES summate more than those of IS, as seen on the difference traces (ES-IS) the effect is greater as frequency increases. At more depolarized potential (-63 mV) the summation and boosting add, resulting in greater excitability.

First, we fitted the IS passive response to hyperpolarizing current pulses (Fig. 1). We then fitted the following parameters: membrane capacitance (Cm), input resistance (RI) and the hyperpolarazing-activated cationic conductance (Ih). Next, keeping the membrane capacitance and Ih values constant, we fitted the ES passive response (Fig. 1). The reduced input resistance of the ES trace was compared to the IS trace accounts for the difference observed in TC between the IS and ES groups.

The differences in the active properties were accounted for by fitting the IS AP data. Using the fitted passive parameters (Cm and RI) and Ih, we fitted the rising part of the AP adjusting the following parameters: Na, K-delayed rectifier (Kdr) and the A-type K channel ($K_A$). Then we proceeded to fit the whole action potential (Fig. 1), fixing the previous parameters, adding the T-type Ca conductance ($Ca_T$), and the non-specific voltage-dependent cationic conductance ($I_{CAN}$). Afterwards small manual parameter adjustments were performed to fine tune the fit.

To fit the ES data we started with the IS parameters and only made changes in as few conductances as possible. This was accomplished by changing two parameters: Na and CaT.

An increase of the Na channel density results in an increase of the AP amplitude and rise rate(Fig. 1). The increase if the CaT conductances greatly increase the ADP. Together, these changes (63% increase on the Na and 185% increase on the $Ca_T$) account for the differences between IS and ES.

Next, we decided to test if the excitability extended to the processing of synaptic inputs by the neurons. We studied synaptic boosting, or the increase in amplitude and duration of the synaptic response as the resting potential of the neuron becomes more depolarized. We delivered the same simulated synaptic inputs (conductance and place in the dendrite) for both the IS and ES conditions at different membrane potentials (Fig. 2). We found that synaptic boosting occurs at more hyperpolarized resting potentials in ES compared to IS, from a 6% increase in synaptic are at -70 mV to a 136% increase at -61 mV (Fig. 2).

When multiple synaptic responses arrive at the soma at a high enough frequency temporal summation increases due to voltage-dependent Na+ channel amplification that boosts the synaptic response and slows the EPSP decay. Ten identical synaptic inputs were delivered in the dendrite at different frequencies, we found that synaptic summation increased the total synaptic area from 9% at 20 Hz to 19% at 50 Hz in ES compared to IS. In addition, the total synaptic maximum amplitude increased from 1.2 mV at 20 Hz to 5.2 mV at 50 Hz (Fig. 2). Both synaptic boosting and synaptic summation results in ES enhanced neuronal output compared to the IS group.

## DISCUSSION

Here we modeled the intrinsic membrane changes we observed in current-clamp recordings from prefrontal cortical pyramidal neurons resulting from control over stress. Patch clamp experiments in our laboratory indicated that rats with escapable control over stress had neurons with faster time constants, increased action potential amplitudes and faster rise rates. These results could be simulated by increasing the leak and Na+ channel conductances. Furthermore, the increased afterdepolarization observed in our recordings could be modeled by increasing the T-type Ca$^{2+}$ conductance in the stressor escape group. Lastly, the model predicts enhanced synaptic boosting and temporal summation resulting from the membrane changes associated with control over stress. Together these result indicate that control over stress increases synaptic throughput and action potential output from prefrontal cortical neurons.

## METHODS

The conductance-based model NEURON (www.neuron.yale.edu) and all model files and expanded methods can be found online at http:// www.neuro-cloud.net/nature-precedings/varela/ .

**PROGRESS AND COLLABORATIONS**

To see up to date progress on this project or if you are interested in contributing to this project visit:
http://www.neuro-cloud.net/nature-precedings/varela/

**ACKNOWLEDGEMENTS**

This work was supported by National Institute on Drug Abuse grant R01-DA24040 (to D.C.C.), NIDA K award K-01DA017750 (to D.C.C.),